\documentclass[cits]{PoS}

\newcommand{\beq}{\begin{equation}}
\newcommand{\eeq}{\end{equation}}
\newcommand{\beqn}{\begin{eqnarray}}
\newcommand{\eeqn}{\end{eqnarray}}

\newcommand{\eps}{\epsilon}

\def\<{\left\langle}
\def\>{\right\rangle}

\def\D{\mathcal{D}}

\def\H{\hat{H}}
\def\M{\hat{M}}
\def\dt{\Delta\tau}
\def\dn{\Delta n}
\def\t{\tau}
\def\a{\hat{a}}
\def\b{\hat{b}}
\def\q{\hat{q}}

\newcommand{\Tr}[1]{ Tr\left( #1 \right) }
\newcommand{\lr}[1]{ \left( #1 \right) }

\title{Monte-Carlo study of the phase transition in the AA-stacked bilayer graphene}

\ShortTitle{Monte-Carlo study of the phase transition in the AA-stacked bilayer graphene}

\author{\speaker{Alexander Nikolaev}\\
        School of Biomedicine, Far Eastern Federal University, Vladivostok, 690950 Russia\\
        E-mail: \email{nikolaev.aa@dvfu.ru}}

\author{Maksim Ulybyshev\\
        Institute for Theoretical Problems of Microphysics, Moscow State University, Moscow, 119899 Russia\\
        Institute of Theoretical and Experimental Physics, Moscow, 117218 Russia\\
        Institute of Theoretical Physics, University of Regensburg, D-93053 Germany, Regensburg, Universit{\"a}tsstra{\ss}e 31\\
        E-mail: \email{ulybyshev@goa.bog.msu.ru}}

\abstract{Tight-binding model of the AA-stacked bilayer graphene with screened electron-electron interactions has been studied using the Hybrid Monte Carlo simulations on the original double-layer hexagonal lattice. Instantaneous screened Coulomb potential is taken into account using Hubbard-Stratonovich transformation. G-type antiferromagnetic ordering has been studied and the phase transition with spontaneous generation of the mass gap has been observed. Dependence of the antiferromagnetic condensate on the on-site electron-electron interaction is examined.}

\FullConference{The 32nd International Symposium on Lattice Field Theory,\\
		23-28 June, 2014\\
		Columbia University New York, NY}

\begin{document}

\section{Introduction}
Graphene is a two-dimensional material, which consist of carbon atoms forming a hexagonal lattice. This material has many unusual electronic and transport properties \cite{Castro_Neto_review}\cite{Rozhkov_review}, which makes it prominent for possible applications in electronics. From the other side graphene may be considered as a desktop laboratory, where one can study nontrivial QFT effects, e.g., atomic collapse, Aharonov-Bohm effect, quantum Hall effect \cite{Katsnelson_book}. Graphene also plays a role of basic element for a broad class of derived materials: bilayer graphene, multilayer graphene, graphane, etc. Different types of bilayer graphene are being studied extensively nowadays, because this material is a good candidate for the basis of electronic devices with a tunable energy gap.

\begin{figure}[t]
\begin{center}
    		\includegraphics[width = 0.75\textwidth]{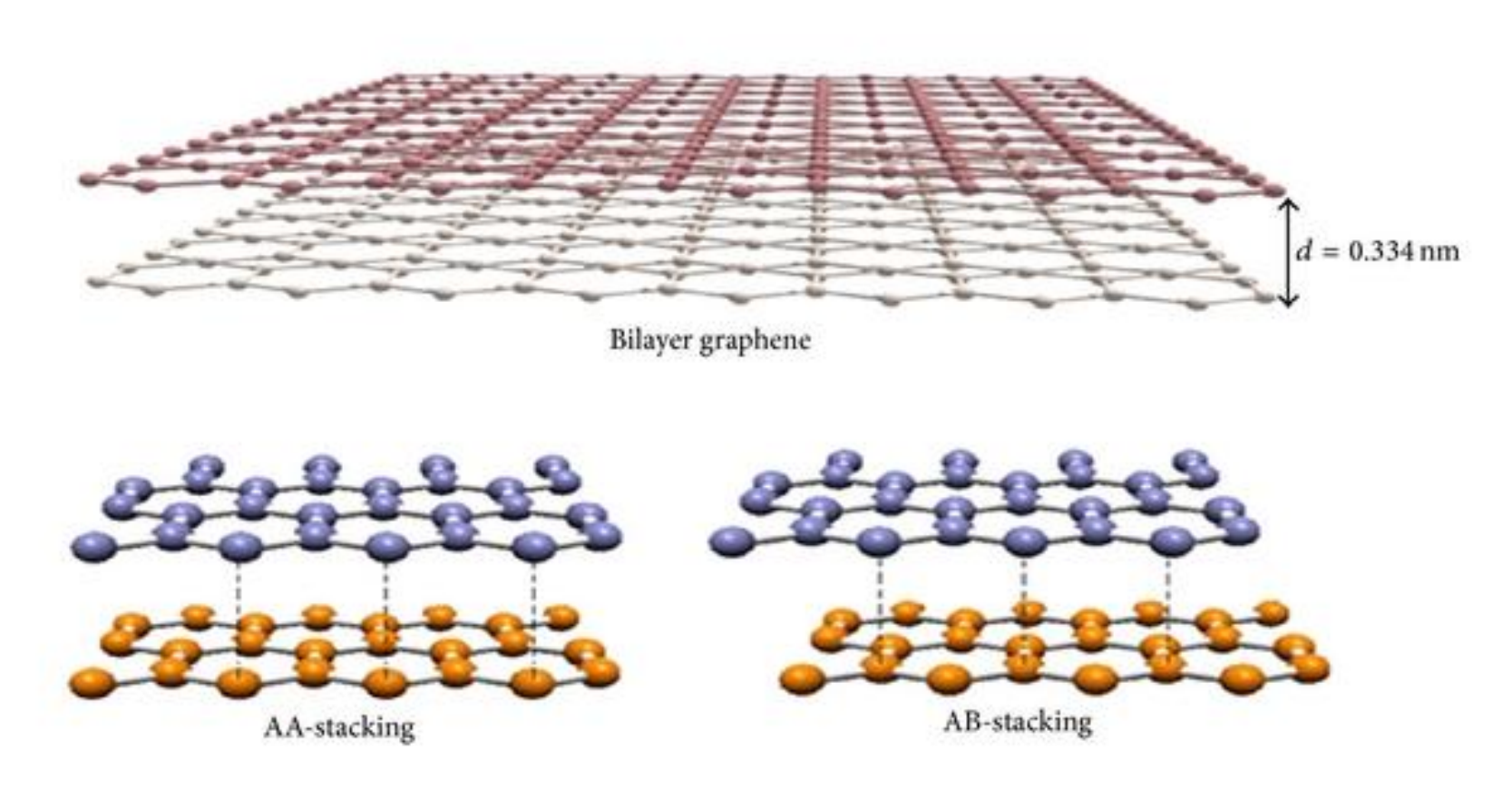}
\caption{Two of the many possible ways of stacking in bilayer graphene.}
 		\label{fig:bilayer_structure}
\end{center}
\end{figure}

The properties of multilayer graphene depend heavily on the way of layer stacking. $AA$-stacked bilayer graphene (AA-BLG) consists of the two graphene layers, stacked in such a way, that each atom of the first layer is located exactly under the corresponding atom of the same sublattice of the second layer (cf. fig. \ref{fig:bilayer_structure}). In recent years AA-BLG have received very limited attention \cite{Borysiuk,Prada_energy_bands,Rozhkov_Vxx,Rozhkov_doping}, probably, because nowadays it is difficult to fabricate high-quality samples of the AA-BLG \cite{Liu_experiment}\cite{Borysiuk}. Energy spectrum of the AA-BLG is linear at low energies, consists of four bands, one electron band and one hole band cross the Fermi energy, and the Fermi surfaces of these two bands coincide \cite{Prada_energy_bands}. Degeneration of the energy spectrum mentioned above makes the system unstable with respect to the formation of interaction-induced energy gap. In particular, on-site Coulomb interaction may lead to the generation of the energy gap accompanied by the formation of G-type AFM ordering (each spin is antiparallel to all of its nearest neighbours) \cite{Rozhkov_Vxx}.

It is possible to generalize the lattice formalism, developed for the monolayer graphene in \cite{Buividovich_Polikarpov} and \cite{Ulybyshev_Katsnelson}, to the case of the AA-BLG. This model allows us to take into account not only on-site electron-electron interaction, but also long-range Coulomb interaction, which may be responsible for some nontrivial physical effects.

\section{Lattice model}
Model Hamiltonian of the AA-BLG has the following form:
\beq\label{H_full}
\H = \H_{tb} + \H_{stag.} + \H_{int.},
\eeq
where $\H_{tb}$ defines the tight-binding part, $\H_{stag.}$ --- "staggered" part and $\H_{int.}$ --- interaction Hamiltonian. Tight-binding Hamiltonian looks like:
\beq\label{H_tb}
\H_{tb} = - t\sum_{i = 1}^2\sum_{<X_i, Y_i>}(\a_{X_i}^+\a_{Y_i} + \b_{X_i}^+\b_{Y_i}) - t_0\sum_{X}(\a_{X_1}^+\a_{X_2} + \b_{X_1}^+\b_{X_2}) + h.c.,
\eeq
where $i$ is the layer index, $X$ and $Y$ are site indices, the sum $\sum_{<X_i, Y_i>}$ is performed over all nearest-neighbour sites in the $i$th layer, $\a_{X_i}^+$, $\a_{X_i}$ and $\b_{X_i}^+$, $\b_{X_i}$ are the creation/annihilation operators for electrons and holes respectively. These operators are connected with standard creation and annihilation operators for the electron with spin up/spin down in the following way:
\beq\label{a_new}
		\a^+_{X,i} = \a^+_{X,i\uparrow},
\eeq
\beq\label{b_new}
 		\b^+_{X,i} = \pm \a_{X,i\downarrow},
\eeq
where the sign in (\ref{b_new}) depends on the layer and sublattice: plus for the case of layer 1, sublattice A and layer 2, sublattice B, minus otherwise. $t$ and $t_0$ in (\ref{H_tb}) represent hopping energies of electrons to the nearest-neighbour within the one layer and to the nearest-neighbour in another layer respectively. Their values ($t = 2.57$ eV, $t_0 = 0.36$ eV) were taken from the paper \cite{Charlier_hoppings}.

Staggered potential is defined as follows:
\beq\label{H_stag}
\H_{stag.} = m\sum_{i = 1}^2\sum_{X, Y}\Bigl [ (-1)^{i+1}\delta_{X_AY_A} + (-1)^{i}\delta_{X_BY_B} \Bigr ](\a_{X_i}^+\a_{Y_i} + \b_{X_i}^+\b_{Y_i}),
\eeq
where $m$ represents bare mass of fermions and $\delta_{X_AY_A}$ ($\delta_{X_BY_B}$) means, that the sites $X$ and $Y$ are equal and both belong to sublattice $A$ ($B$). We need this "staggered" potential for two reasons: firstly, it serves as a primer for the ground state of G-type AFM ordering, for which the formation of energy gap was predicted, and secondly, it regularizes zero modes of the Dirac operator. In the absence of interaction AA-BLG has no energy gap \cite{Prada_energy_bands}, thus the lattice calculations have been performed for a few values of $m$ and than observables were extrapolated to the point $m$ = 0.

Finally, Coulomb part of the Hamiltionian (\ref{H_full}) has the form:
\beq\label{H_Coulomb}
\H_{int.} = \frac{1}{2}\sum_{i,j = 1}^2\sum_{X,Y}\q_{X_i} V_{XY}^{ij} \q_{Y_j},
\eeq
where $\q_{X_i} = \a^+_{X_i}\a_{X_i} - \b^+_{X_i}\b_{X_i}$ is the charge operator for the site $X$ on the layer $i$ and $V_{XY}^{ij}$ represents interaction potential between the sites $X_i$ and $Y_j$ (because the Fermi velocity is approximately $c/315$ in AA-BLG, electron-electron interaction may be considered as instantaneous). The introduced potential $V_{XY}^{ij}$ takes into account the screening by electrons at $\sigma$-orbitals at small distances, it is piecewise-defined: within one layer the values for the on-site interaction potential ($V_{xx}$), the potentials between nearest neighbours ($V_{01}$), next-to-nearest neighbours ($V_{02}$) and next-to-next-to-nearest-neighbours ($V_{03}$) are fixed and taken from \cite{Wehling} (table I, 3d column), while the potential at distances $r > 2a$ is considered as Coulomb, starting from $V_{03}$. Potential for the sites, located at different layers, is introduced as usual Coulomb potential. Taking into account the screening effect in lattice model of monolayer graphene and AA-BLG can be crucial point for reproduction of real physics in simulations, because effects of electron-electron interaction are very sensitive to the modification of the short-range potentials \cite{Ulybyshev_Katsnelson}.

Hereafter we will use the standard formalism of statistical QFT \cite{Montvay}\cite{Gattringer}. The partition function for the system is defined as follows:
\beq\label{Partition_function}
Z = \Tr{e^{-\beta\H}},
\eeq
where $\beta = 1/T$ is the inverse temperature of electron gas. To calculate the partition function we have to perform Suzuki-Trotter decomposition:
\beqn\label{Suzuki_Trotter}
\Tr{e^{-\beta\H}} = \Tr{e^{-\dt(\H_{tb} + \H_{stag.} + \H_{int.})}}^{N_t} = \nonumber\\
    		= \Tr{e^{- \dt(\H_{tb} + \H_{stag.})}e^{- \dt\H_{int.}}e^{- \dt(\H_{tb} + \H_{stag.})}e^{- \dt\H_{int.}} \ldots} + O(\dt^2),
\eeqn
where $\dt = \beta/N_\t$, and insert Grassmannian coherent states between the exponentials \cite{Montvay}. The point here is that $\H_{int.}$ is not the quadratic form of $\a$ and $\b$, so it is convenient to decompose $e^{- \dt\H_{int.}}$ using the Hubbard-Stratonovich transformation \cite{Rebbi}:
\beq\label{Hubbard_Stratonovich}
e^{ - \frac{\dt}{2} \sum\limits_{X,Y}\q_{X}V_{XY}^{ij}\q_{Y}}
= \int\D\varphi e^{ - \frac{1}{2\dt} \sum\limits_{X,Y} \varphi_X V^{-1}_{XY} \varphi_Y - i \sum\limits_{X} \varphi_X \q_X },
\eeq
where $\varphi_X$ defines real-valued scalar field (Hubbard field) at the site $X$ (layer indices on the r.h.s. of (\ref{Hubbard_Stratonovich}) were omitted for notation compactness). After some algebra, we arrive at the following expression for the partition function:
\beq\label{Partition_final}
Z = \int \D \varphi \textsl{det}(M^+M) e^{- S[\varphi]},
\eeq	
where $M$ is the fermionic operator for electrons, 
\beq\label{Hubbard_action}
S[\varphi] = \frac{1}{2\dt} \sum_{\t} \sum_{i,j = 1}^2 \sum_{X,Y} \varphi_{Xi}^{\t} \lr{\hat{V}^{-1}}^{ij}_{XY} \varphi_{Yj}^{\t}
\eeq
represents Hubbard action, and Hubbard fields in this sum are located only at odd time layers. It is important to note, that due to the decomposition (\ref{Suzuki_Trotter}) the space-time lattice has $2N_t$ time layers, p.b.c. in time direction are introduced for Hubbard fields. Dirac operator $M$ has the following form ($t = 0,\ldots,N_t - 1$, a.p.b.c. in time direction are considered for fermion fields):
\beq\label{M_formula}
M_{XiYj}^{\tau'\tau} = \delta^{\tau', \tau}\delta_{ij}\delta_{XY} - \delta^{\tau', 2t}(\delta_{ij}\delta_{XY} + \Delta\tau A_{XiYj})\delta^{\tau, 2t + 1} - \delta^{\tau', 2t + 1}\delta_{ij}\delta_{XY}e^{i\varphi_{Xi}^{2t +2}}\delta^{\tau, 2t + 2}
\eeq
where $A^{XiYj}$ is a real matrix and is defined as follows:
\beqn\label{A_matrix}
A_{XiYj} = t \delta_{ij} \lr{ \delta_{X \in A}\sum_{b=0}^2 \delta_{Y , X + \rho_b} + \delta_{Y \in B}\sum_{b=0}^2 \delta_{X , Y - \rho_b} } + t_0 \delta_{XY} \lr{ \delta_{i1}\delta_{j2} + \delta_{i2}\delta_{j1} } -\\  - m \delta_{ij} \lr{ \lr{-1}^{i+1}\delta_{X_AY_A} + \lr{-1}^i\delta_{X_BY_B} }.\nonumber
\eeqn
$\rho_b$ in (\ref{A_matrix}) denote a set of three vectors, which point from the site of the sublattice $A$ within hexagonal lattice to its nearest neighbours belonging to sublattice $B$. 

The formation of the energy gap in the AA-BLG energy spectrum may be examined by the calculation of the AFM condensate:
\beq\label{d_n_AB}
\dn = n_{1A\uparrow} - n_{2A\uparrow} = n_{1B\downarrow} - n_{2B\downarrow},
\eeq
where $n_{i\alpha\uparrow(\downarrow)}$ denotes the number of electrons with spin up (spin down) on the layer $i = 1, 2$ and sublattice $\alpha = A, B$. Only simultaneous breaking of interlayer and sublattice symmetries may lead to the gap opening \cite{Rozhkov_Vxx}. Moreover, the energy gap may be evaluated within the mean field approximation as $\Delta = V_{xx}\dn/2$ at small enough $\dn$. Thermodynamic expectation variable for (\ref{d_n_AB}) has the following form:
\beq\label{AFM_condensate}
\< \dn \> = \frac{1}{N_\t N}\sum_{\t}\< \sum_{X \in A} \Bigl( \M^{-1}_{X2X2} - \M^{-1}_{X1X1} \Bigr) \> = \frac{1}{N_\t N}\sum_{\t}\< \sum_{X \in B} \Bigl( \M^{-1}_{X1X1} - \M^{-1}_{X2X2} \Bigr) \>,
\eeq
where $N$ represents the number of spatial lattice sites in one sublattice.

\begin{figure}[t]
\begin{center}
	\begin{minipage}[t]{0.49\textwidth}
    		\includegraphics[width = 1.0\textwidth]{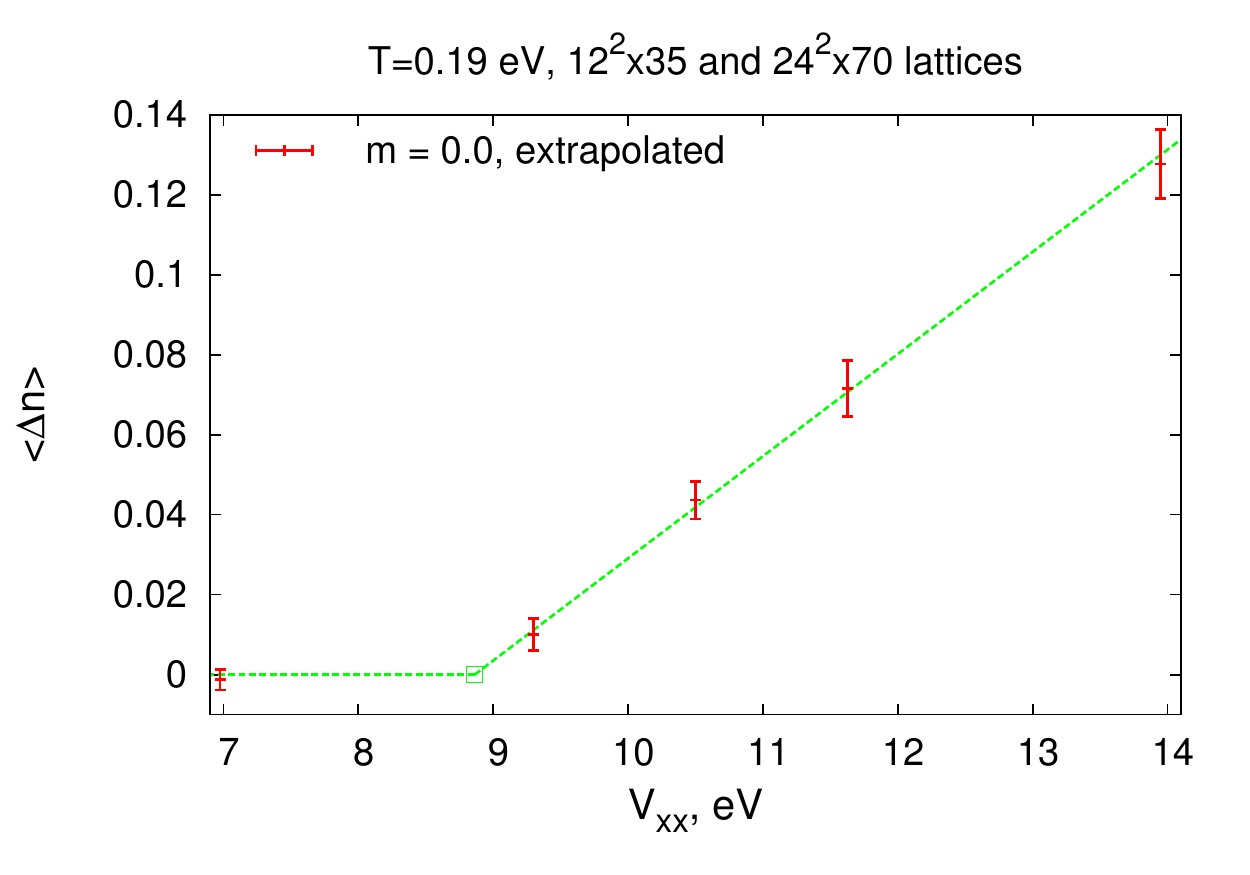}
\caption{The dependence of the AFM condensate on the on-site interaction potential $V_{xx}$ at fixed temperature. AFM condensate vanishes at the value $V_{xx}^c = (8.89 \pm 0.33)$ eV.}
 		\label{fig:Vxx_dependence}
    \end{minipage}
\hfill
    \begin{minipage}[t]{0.49\textwidth}
    		\includegraphics[width = 1.0\textwidth]{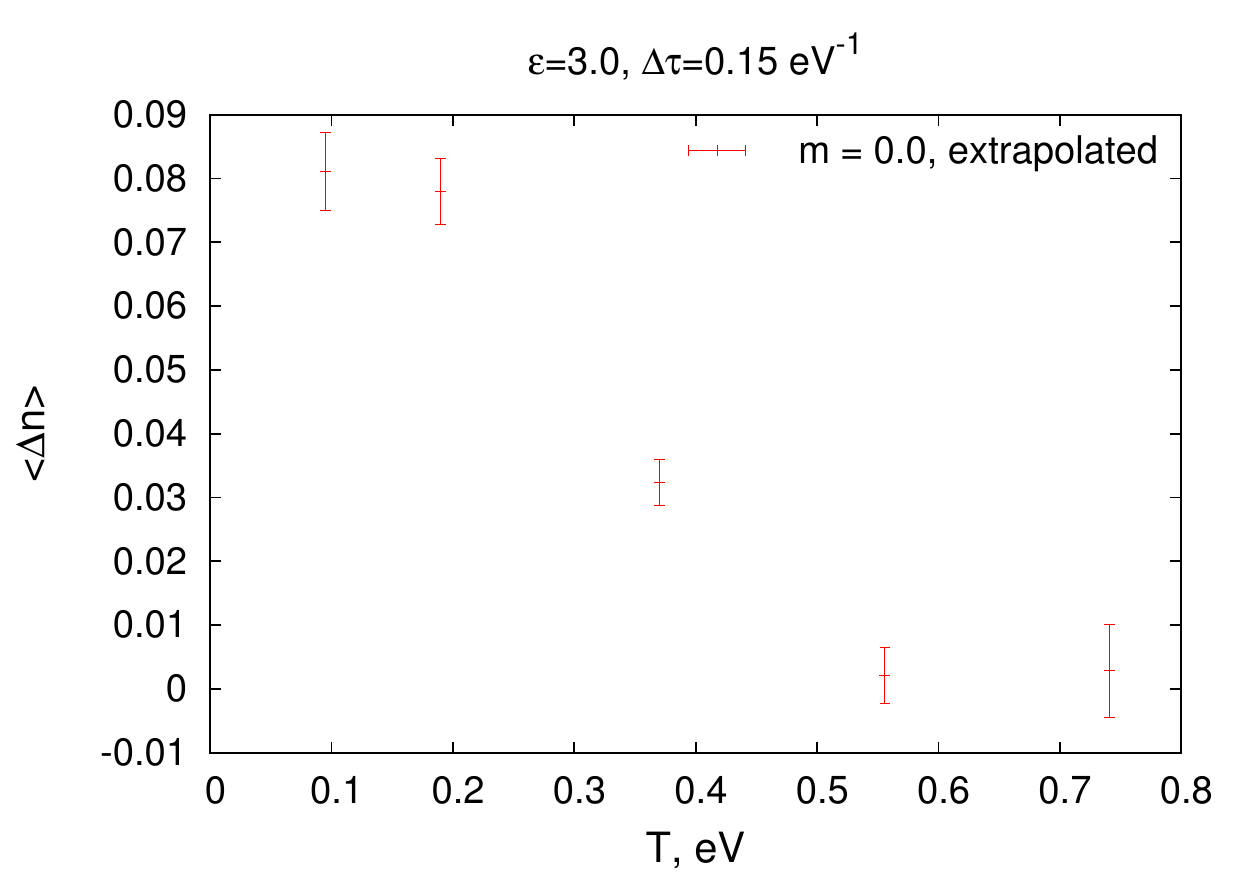}
\caption{The dependence of the AFM condensate on the temperature. All potentials, except $V_{xx}$, were rescaled by the factor of $\eps = 3.0$: $V^{ij}_{XY} \rightarrow V^{ij}_{XY}/3.0$.}
 		\label{fig:temperature_dependence}
    \end{minipage}
\end{center}
\end{figure}

\section{Numerical results and discussion}
Simulations were performed on the lattices with spatial size from $12\times12$ to $24\times24$ and temporal sizes from $N_\t = 9$ to $N_\t = 70$, $\dt = 0.15$ eV$^{-1}$ (the temperature of electron gas is defined as $T = 1/(N_\t\dt)$, which follows from (\ref{Suzuki_Trotter})). The following bare fermion masses were examined for each lattice: $m = 0.1$ eV, $0.15$ eV, $0.2$ eV, $0.25$ eV, $0.3$ eV; than the quadratic extrapolation to $m = 0.0$ was performed for the calculation of the AFM condensate. We generated 150 --- 300 statistically independent configurations for every value of mass, $\Phi$-algorithm was used in simulations.

The first result is presented at Fig. \ref{fig:Vxx_dependence} and shows the dependence of AFM condensate (\ref{AFM_condensate}) on the on-site interaction potential $V_{xx}$ (other potentials are fixed) at particular temperature. This dependence is almost linear and linear extrapolation yields to the critical value $V_{xx}^c = (8.89 \pm 0.33)$ eV, where $\< \dn \>$ vanishes. On the contrary, mean-field approximation with only on-site Coulomb interaction taken into account predicts nonvanishing exponential dependence for the energy gap \cite{Rozhkov_Vxx}. From this one can conclude, that long-range electron-electron interactions work against the formation of the energy gap and produce a noticeable impact on the collective phenomena in the AA-BLG.

The second result is shown at Fig. \ref{fig:temperature_dependence} and represents the dependence of $\< \dn \>$ on temperature for the fixed values of potentials. In these measurements all potentials, except $V_{xx}$, were rescaled by the factor of $\eps$: $V^{ij}_{XY} \rightarrow V^{ij}_{XY}/\eps$ (inscription "$\eps = 3.0$" in the caption over the graph). Thus $\eps$ may be considered as a dielectric permittivity of the substrate, on which AA-BLG sample is located. The plateua for $\< \dn \>$ at small temperatures and the transition at $T \approx 0.37$ eV can be clearly seen at the Fig. \ref{fig:temperature_dependence}, and $\< \dn \>$ can be considered as an order parameter for the AFM phase transition. At high temperatures $\< \dn \> \approx 0$, what means, that AFM ordering is destroyed by thermal fluctuations. 

\begin{acknowledgments}
 The authors are grateful to Prof. Oleg Pavlovsky for interesting and useful discussions. The work was supported by the Grant RFBR-14-02-01261-a and by the Scientific Fund of FEFU, under Contract No. 12-02-13000-m-02/13. MU acknowledges support from the Deutsche Forschungsgemeinschaft grant SFB/TR-55. Numerical calculations were performed at the ITEP systems Graphyn and Stakan (authors are much obliged to A.V. Barylov, A.A. Golubev, V.A. Kolosov, I.E. Korolko, M.M. Sokolov for the help) and at Supercomputing Center of the Moscow State University.
\end{acknowledgments}


\begin{thebibliography}{99}
\bibitem{Castro_Neto_review}
A.~H.~Castro Neto, F.~Guinea, N.~M.~R.~Peres, K.~S.~Novoselov, and A.~K.~Geim, Rev. Mod. Phys. \textbf{81}, 109--162 (2009).

\bibitem{Rozhkov_review}
A.~V.~Rozhkov, G.~Giavaras, Y.~P.~Bliokh, V.~Freilikher, and F.~Nori, Physics Reports \textbf{503}, 77--114 (2011).

\bibitem{Katsnelson_book}
M.~I.~Katsnelson, \emph{Graphene: Carbon in Two Dimensions}, Cambridge University Press, Cambridge 2012.

\bibitem{Borysiuk}
J.~Borysiuk, J.~Soltys, and J.~Piechota, Journal of Applied Physics \textbf{109}, 093523 (2011).

\bibitem{Prada_energy_bands}
E.~Prada, P.~San-Jose, L.~Brey, H.~A.~Fertig, Solid State Communications \textbf{151},
1075--1083 (2011).

\bibitem{Rozhkov_Vxx}
A.~L.~Rakhmanov, A.~V.~Rozhkov, A.~O.~Sboychakov, and F.~Nori, Phys. Rev. Lett. \textbf{109}, 206801 (2012).

\bibitem{Rozhkov_doping}
A.~O.~Sboychakov, A.~L.~Rakhmanov, A.~V.~Rozhkov, and F.~Nori, Phys. Rev. B \textbf{87}, 121401 (2013).

\bibitem{Liu_experiment}
Z.~Liu, K.~Suenaga, P.~J.~F.~Harris, and S.~Iijima, Phys. Rev. Lett. \textbf{102}, 015501 (2009).

\bibitem{Buividovich_Polikarpov}
P.~V.~Buividovich and M.~I.~Polikarpov,
  Phys. Rev. B \textbf{86}, 245117 (2012)
 
\bibitem{Ulybyshev_Katsnelson}
M.~V.~Ulybyshev, P.~V.~Buividovich, M.~I.~Katsnelson, and M.~I.~Polikarpov, Phys. Rev. Lett. \textbf{111}, 056801 (2013).

\bibitem{Charlier_hoppings}
J.-C.~Charlier, J.-P.~Michenaud, and X.~Gonze, Phys. Rev. B \textbf{46}, 4531--4539 (1992).

\bibitem{Wehling}
T.~O.~Wehling, E.~\ifmmode \mbox{\c{S}}\else \c{S}\fi{}a\ifmmode \mbox{\c{s}}\else \c{s}\fi{}\ifmmode \imath \else \i \fi{}o\ifmmode \breve{g}\else \u{g}\fi{}lu, C.~Friedrich, A.~I.~Lichtenstein, M.~I.~Katsnelson, and S.~Bl\"ugel, Phys. Rev. Lett. \textbf{106}, 236805 (2011).

\bibitem{Montvay}
I.~Montvay and G.~M\"{u}nster, \emph{Quantum fields on a lattice}, Cambridge University Press, 
Cambridge 1994.

\bibitem{Gattringer}
C.~Gattringer and C.~B.~Lang, \emph{Quantum Chromodynamics on the Lattice: An Introductory Presentation}, Springer, Berlin Heidelberg 2010.

\bibitem{Rebbi}
R.~C.~Brower, C.~Rebbi, and D.~Schaich, in proceedings of \emph{Lattice 2011}, \pos{PoS(Lattice 2011)056}.

\end{thebibliography}
\end{document}